\documentstyle{proceeding}
\input psfig.tex
\newcommand{\gapr}{\raisebox{-.6ex}{\mbox{
$\stackrel{>}{\mbox{\scriptsize$\sim$}}\:$}}}
\newcommand{\lapr}{\raisebox{-.6ex}{\mbox{
$\stackrel{<}{\mbox{\scriptsize$\sim$}}\:$}}}
\def\R{ROSAT~}
\begin{document}

\title{X-ray spectra from convective photospheres of neutron stars}

\author{V.E. Zavlin\inst{1}, G.G. Pavlov\inst{2,3},
Yu.A. Shibanov\inst{3},
F.J. Rogers\inst{4} \and C.A. Iglesias\inst{4}}

\institute{Max--Planck--Institut f\"ur Extraterrestrische Physik,
 Giessenbachstrasse, D-85740 Garching, Germany
\and Pennsylvania State University, 525 Davey Lab, PA 16802, USA
\and Ioffe Institute of Physics and Technology, 194021,
St.~Petersburg, Russia
\and Lawrence Livermore National Laboratory, Livermore, CA 94550, USA}
\maketitle

\begin{abstract}
We present first results of modeling convective photospheres
of neutron stars. We show that in
photospheres composed of the light elements convection arises only at
relatively low effective temperatures ($\lapr 3-5\times 10^4$ K),
whereas in the case of iron compositon it
arises at $T_{\rm eff}\lapr 3\times 10^5$ K.
Convection changes the depth dependence of the photosphere temperature
and the shapes of the emergent spectra.
Thus, it should be taken into account for the proper interpretation of
EUV/soft-X-ray observations of the thermal radiation from neutron stars.
\end{abstract}

\section{Introduction}

Recent \R observations of pulsars revealed that some of them
emit thermal-like radiation in the soft X-ray range (\"Ogelman 1995). These
observations stimulated further
investigations of neutron star (NS) photospheres responsible for the
properties of the emitted radiation.

One of important and hitherto untouched problems
associated with modeling of the  NS photospheres is the problem of
convective energy transport which can affect the temperature
distribution and the emergent spectra.
Due to huge gravitational accelerations, $\sim 10^{14} - 10^{15}$
cm s$^{-2}$, the NS photospheres are much denser, $\rho \sim 0.001 - 1$
g cm$^{-3}$, than those of usual stars.
The increased densities shift ionization equilibrium: the nonionized
fraction grows with $\rho$ at moderate densities,
$\rho \lapr 0.01-0.1$ g cm$^{-3}$,
and sharply decreases at $\rho \gapr 0.1- 1$ g cm$^{-3}$ due to pressure
ionization. As a result, zones of increased opacity
(increased radiative gradient $\nabla_{\rm rad}$) and reduced adiabatic
gradient $\nabla_{\rm ad}$  develop in the NS photospheres at
temperatures much higher
than  in photospheres of usual stars, which may cause convective
instability at  depths where the emergent spectrum is formed.
So far the NS photospheres have
been considered  either for  stars with very strong surface magnetic fields
$B\sim 10^{12} - 10^{13}$ G (see, for example, Pavlov $et~al$. 1994)
where existing convection theories are not applicable, or for `nonmagnetic'
(low-field) photospheres, $B < 10^8 - 10^9$ G, with high surface effective
temperatures
$T_{\rm eff} > 10^{5.5} - 10^{6}$ K (Romani 1987), where the convection
can hardly
be expected. Since thermal NS radiation can be detected for $T_{\rm eff}
\gapr 2\times 10^5$ K in the soft X-ray range, and for even lower temeratures,
$\gapr 2\times 10^4$ K,
in the UV/optical range (Pavlov $et~al$. 1995), the study of the convective
NS photospheres is important for the proper interpretation of these
observations.

The convective flow in stellar photospheres is turbulent and imposes many
complicated problems. A common practice, which we follow here, is to use
the phenomenological mixing-length theory with
the traditional Schwarzschild criterion for convective
instability, $\nabla_{\rm rad} > \nabla_{\rm ad}$.
This theory has been widely implemented and testified.

Details of our numerical calculations will be described elsewhere.
Generally, we employ the complete linearization method for computing
the photosphere models and include the convective energy transfer as
described by Mihalas (1978).

\section{Results and Discussions}

We calculated the model of nonmagnetic photospheres for different
chemical compositions.
Here we present examples for pure hydrogen, helium and iron
compositions
at the gravitational acceleration $g=2.43
\times 10^{14}$ cm s$^{-2}$ which corresponds to standard NS mass
$M=1.4 M_\odot$ and radius $R=10$ km. The raditive opacities and
equation of state for the iron composition were taken from the
OPAL library (Iglesias $et~al$. 1992). The results can be directly
applicable to very old NSs with low magnetic fields
(e.~g., millisecond pulsars).

Our results show that, similar to the case of usual stellar
photospheres,
the convective energy transfer begins to play a role at lower surface
temperatures when  atoms are not fully ionized and
the radiative opacities are strongly increased by
contribution of the bound-free and bound-bound transitions.
The increased opacities result
in high values of $\nabla_{\rm rad}$
--- the depth dependence of the temperature
becomes steeper in order to transfer the energy flux throughout
the photosphere. On the other hand,
$\nabla_{\rm ad}$ in the dense partially ionized layers can be
much smaller than its limiting value 0.4 for an ideal fully ionized
or fully nonionized gas (see, for example, Cox and Guili 1968).
As a result, superficial convection zones in NS photospheres
 arise and,
as in usual stars, they are associated
with layers of partial ionization.
Our computations show that $\nabla_{\rm ad}$ can drop down
to $\simeq 0.1$ when the nonionized fractions are $\gapr$70\%.

The actual temperature gradient $\nabla$ in the convection zones
satisfies the following relation: $\nabla_{\rm ad} <
\nabla < \nabla_{\rm rad}$. Fig. 1 shows the
temperature distributions
in photospheres with and without allowance for convection.
One can see that convection leads to more gradual profiles,
in accordance with the above relation.
Both in very surface layers, where  $\nabla_{\rm rad}$
is too low, and in deep layers, where $\nabla_{\rm ad}$
is close to its maximum value, convection is absent,
and the temperature profiles remain the same.

\begin{figure}
\vspace{5.5cm}
\caption[]{Temperature profiles with and without allowance for
convection (solid and dashed curves, respectively) for H,
He and Fe photospheres
($T_{\rm eff}=1.8\times 10^4,~4.1\times 10^4$
and $1.4\times 10^5$ K, respectively).}
\end{figure}

The convective transfer affects not only the structure of photosphere but
also the spectra of the NS thermal radiation
(Fig. 2) because the temperature profiles are
changed in the layers where the radiation escapes from.
In particular, convection substantially (up to two orders of
magnitide) lowers the flux from H and He photospheres at photon
energies above the main photoionization edges, so that the
high-energy spectral tails
become softer. The spectra remain the same at low and very high
energies since both  shallow and very deep layers are not
affected by convection. In the case of Fe composition, the convective zone
lies so deep that only a high-energy tail ($E\gapr 0.3$ keV) is affected.
The effect of convection on the spectra
disappears with increasing effective temperature
({\it e.~g.}, at $T_{\rm eff}>3\times 10^4$ K for H, at
$T_{\rm eff}>5\times 10^4$ K for He and
at $T_{\rm eff}>3\times 10^5$ K for Fe
photospheres).

\begin{figure}
\vspace{8.4cm}
\caption[]{Spectra of outgoing radiation corresponding to the temperature
profiles in Fig. 1. Dash-dotted curves are the blackbody spectra
$\pi B(T_{\rm eff})$.}
\end{figure}

The presented results correspond to the convective efficiency
$l/H=1$. Acceptable values of this parameter can
vary from 0.3 to 2.5.
Bergeron $et~al$. (1992) showed
that higher efficiency enhances convection and smoothes the spectra
of white dwarfs.
The same effect can be expected in the NS
photospheres and, consequently, the convection there may develop at
higher effective temperatures than in the models presented.

Convection in NS photospheres is important because it can mix
the material in convective zones, bringing heavier
elements from  bottom to surface layers which would
otherwise contain only light elements due to gravitational
stratification. Our calculations show that this may happen
in cold NSs with $T_{\rm eff}\lapr (3-5)\times 10^4$ K,
whereas radiating layers of hotter NSs can be expected to consist
mainly of hydrogen and helium. Since the presence of convection
softens the high-energy tails of the spectra,
this effect should be taken into account for the proper interpretation
of EUV/soft-X-ray observations of thermal radiation from NSs.

\vskip 0.4cm
\begin{acknowledgements}
This work was partly suppoted
by INTAS  grant 94-3834, NASA grant NAG5-2807 and
RFFI grant 93-02-2916.
\end{acknowledgements}

\end{document}